

\input harvmac.tex
\noblackbox
\def\lae{\raise-.5ex\vbox{\hbox{$\; <\;$}\vskip-2.9ex\hbox{$\; \sim\;$}}}
\def\gae{\raise-.5ex\vbox{\hbox{$\; >\;$}\vskip-2.9ex\hbox{$\; \sim\;$}}}
\def\slash#1{\raise.15ex\hbox{/}\kern-.57em #1}

\def\ie{{\it i.e.}}

\def\two#1{\raise1.35ex\hbox{$\leftrightarrow$}\kern-.88em#1}
\def\lefta#1{\raise1.35ex\hbox{$\leftarrow$}\kern-.61em#1}
\def\righta#1{\raise1.35ex\hbox{$\rightarrow$}\kern-.61em#1}
\def\Dslash{\raise.15ex\hbox{/}\kern-.77em D}
\def\np#1#2#3{Nucl. Phys. {\bf #1} (#2) #3}
\def\pl#1#2#3{Phys. Lett. {\bf #1} (#2) #3}
\def\prl#1#2#3{Phys. Rev. Lett. {\bf #1} (#2) #3}
\def\pr#1#2#3{Phys. Rev. {\bf #1} (#2) #3}
\def\prd#1#2#3{Phys. Rev. D {\bf #1} (#2) #3}
\def\vsl{\raise.15ex\hbox{/}\kern-.57em v}
\def\ie{{\it i.e.}}

\def\etal{{\it et al.}}

\Title{\vbox{\hbox{BUHEP--92--25}\hbox{HUTP--92/A033}\hbox{hep-ph/9207249}%
}}{ \vbox{\centerline{Electroweak Corrections in Technicolor Reconsidered}}}


\centerline{R. Sekhar Chivukula$^{a,1}$}
\medskip\centerline{Michael J. Dugan$^{a,b,2}$}
\medskip\centerline{\it and}
\medskip\centerline{Mitchell Golden$^{a,3,*}$}
\footnote{}{$^a$Boston University, Department of Physics, 590 Commonwealth
Avenue, Boston, MA 02215}
\footnote{}{$^b$Lyman Laboratory of Physics,
Harvard University, Cambridge, MA 02138}
\footnote{}{$^1$sekhar@weyl.bu.edu, $^2$dugan@huhepl.bitnet,
$^3$golden@weyl.bu.edu}
\footnote{}{$^*$Address after Sept 1, 1992: Lyman Laboratory of Physics,
Harvard University, Cambridge, MA 02138}

\vskip .1in

Radiative corrections to electroweak parameters in technicolor theories may be
evaluated by one of two techniques: either one estimates spectral function
integrals using scaled QCD data, or one uses naive dimensional analysis
with a chiral Lagrangian. The former yields corrections to electroweak
parameters proportional to the number of flavors and the number of colors,
while the latter is proportional to the number of flavors squared and is
independent of the number of colors. We attempt to resolve this apparent
contradiction by showing that the spectrum of technicolor one obtains by
scaling QCD data to high energies is unlikely to resemble that of an actual
technicolor theory. The resonances are likely to be much lighter than
naively supposed and the radiative corrections to electroweak parameters
may by much larger. We also argue that much less is known about the
spectrum and the radiative corrections in technicolor than was previously
believed.

\Date{7/92} 

\newsec{Introduction}

Because of the increased precision with which quantities like the $W$ and
$Z$ masses are known, there has recently been a great deal of interest in
constraining new physics by its radiative effects. The simplest kinds of
radiative corrections are ``oblique'' \ref\LPS{B. Lynn, M. Peskin, and R.
Stuart, in {\it Trieste Electroweak 1985} Proceedings.}, meaning that they
affect only the propagator of the electroweak gauge bosons. If the new
physics is heavy enough, then the radiative effects are quantified in three
parameters \ref\PTI{M.~Peskin and T.~Takeuchi, \prl{65}{1990}{964}.},
commonly referred to as $S$, $T$, and $U$. Much work has focused on the
evaluation of these quantities in technicolor theories \ref\techni{S.
Weinberg, \prd {19} {1979} {1277}\semi L. Susskind,
\prd {20} {1979} {2619}.}.

Since technicolor is a strongly interacting theory, the corrections cannot
be evaluated by ordinary perturbation theory. One of two approaches is
usually followed. The first involves expressing $S$, for example, as a
spectral integral, which is then evaluated by taking data from QCD
experiments and scaling it to technicolor energies
\PTI\ref\spect{R.~N.~Cahn and M.~Suzuki, \prd{44}{1991}{3641}.}
\ref\PTII{M.~Peskin and T.~Takeuchi, \prd{46}{1992}{381}.}.
The second is to use chiral Lagrangian techniques
\ref\CCWZ{S. Weinberg, \pr {166} {1968}{1568}\semi
S. Coleman, J. Wess, and B. Zumino, \pr {177} {1969} {2239} \semi
C.  Callan, S. Coleman, J. Wess, and B. Zumino \pr {177} {1969} {2246}.}.
In the latter case $S$, $T$, and $U$ are related to coefficients of
four-derivative operators in a chiral Lagrangian
\ref\longh{A. Longhitano, \prd{22}{1980}{1166} and \np {B188}{1981}{118}.}
\ref\RP{R.~Renken and M.~Peskin, \np{B211}{1983}{93}.}
\ref\GR{M.~Golden and L.~Randall, \np{B361}{1990}{3}.}
\ref\HT{B.~Holdom and J.~Terning, \pl{B247}{1990}{88}.}
\ref\DEH{A.~Dobado, D.~Espriu, and M.~J.~Herrero, \pl{B255}{1990}{405}.}
\ref\Grgi{H.~Georgi, \np{B363}{1991}{301}.}.
In general, the coefficients of a chiral Lagrangian are arbitrary, but they
may be estimated by the technique of naive dimensional analysis (NDA)
\ref\NDA{S. Weinberg, Physica {\bf 96A} (1979) 327\semi see also H. Georgi and
A. Manohar, \np {B234} {1984} {189} and
H.~Georgi and L.~Randall, \np{B276}{1986}{241}.}
which states that, roughly speaking, the size of a four-derivative term in the
chiral Lagrangian is set by the typical size of the PGB loops that contribute
to it.

When applied to the one-family model of technicolor \ref\onefamily{E. Farhi
and L.  Susskind, \prd {20} {1979} {3404}.}, both these techniques yield
contributions to the radiative parameters of about the same size.  Yet there
is something a bit odd about this concurrence of the results.  The two
techniques appear to depend differently on the numbers of colors and families
involved.  As we will explain below, the spectral function technique, as
normally implemented, yields values of $S$ proportional\foot{A third approach
to evaluating $S$ in technicolor theories is based on a direct estimate of the
technihadronic contribution to the $W$ and $Z$ vacuum polarization diagrams
\ref\appls{T. Appelquist and G. Triantaphyllou, \pl {B278} {1992} {345}.}
\ref\hsu{R. Sundrum and S. Hsu, ``Walking Technicolor and Electroweak
Radiative Corrections'', LBL preprint LBL-31066, UCB-PTH-91/34 (June 1992).}.
This approach also produces a value of $S$ proportional to $N_D N_{TC}$.} to
$N_D N_{TC}$, where $N_D$ is the number of doublets (four in the one-family
model), and $N_{TC}$ is the number of technicolors in the underlying gauge
theory.  The chiral Lagrangian with NDA, on the other hand, certainly knows
nothing about the number of colors in the underlying theory.  We will show
below that the corrections grow like $N N_D$, where $N$ is the total number of
technifermion flavors.

This note is an attempt to reconcile these two pictures\foot{Ref \PTII\
also addresses this issue. However, their point of view is rather
different from ours.}. We will argue that the scaling of QCD data which one
does in deriving $S$ is based on an unwarranted assumption about the spectrum
of resonance masses in technicolor theories. In particular, we argue that
naively scaling the masses of the resonances in QCD will underestimate the
masses of mesons in technicolor. If this is so, the radiative corrections may
be considerably larger than given by the spectral function estima<tes.  We
will argue that the $N N_D$ dependence is probably applicable when $N_{TC}$ is
small, while $N_D N_{TC}$ holds only for rather large values of $N_{TC}$.

In any case, we argue that much less is known about the spectrum of
technicolor than was previously believed. Though the most pessimistic
evaluations of the viability of the simplest technicolor models may in the
end be justified, more uncertainty remains in the evaluation of the
radiative corrections than was appreciated.

The plan of this paper is as follows.  In the next section we explain in
detail how the factors of $N_D$ and $N_{TC}$ come about in the different
techniques.  The following section we explain an idea suggested by Kaplan
\ref\Kaplan{D.~Kaplan, private communication} which points the way to
resolving the conflict.  The following section applies the argument to
technicolor theories.  In section five we apply these considerations in the
limit of QCD in which both $N$ and $N_{TC}$ go to infinity simultaneously,
and make some concluding remarks in section six.

\newsec{The Two Calculations}

The quantity $S$ is defined by\foot{This discussion of the spectral integral
computation follows ref \PTII.}
\eqn\SdefI{
S = -16 \pi {\partial \over \partial q^2} \Pi_{3 Y}(q^2)|_0~,
}
where $\Pi_{3 Y}$ is the transverse part of the
weak-$T_3$-current-weak-hypercharge-current two-point correlator.  In this
paper we will consider only the case of technicolor theories in which the
symmetry breaking pattern is $SU(N)_L \times SU(N)_R \times U(1) \to SU(N)_V
\times U(1)$.  We embed the weak $SU(2)_W$ into $SU(N)_L$ as $N_D$ doublets and
$N_S=N-2N_D$ singlets.  This embedding preserves a custodial $SU(2)_C$
symmetry.  In this case, we may rewrite $S$ as
\eqn\SdefII{
S = - 4\pi {\partial \over \partial q^2} (\Pi_{VV}(q^2)-\Pi_{AA}(q^2))|_0~,
}
where $\Pi_{VV}$ ($\Pi_{AA}$) is the transverse part of the two-point
correlator of the third component of the weak vector (axial) current.
Defined this way $S$ is infinite; however by subtracting the same
expression in the standard one-Higgs model (with some fixed value of the
Higgs boson mass) one may define a finite quantity which parameterizes the
radiative effects.

The Fermi constant is given by
\eqn\Gf{
{G_F \over \sqrt{2}} = {1 \over 2 \Pi_{AA}(0)}
}
so we identify $\Pi_{AA}(0) = v^2 = \hbox{(246 GeV)}^2$.  Because of the
embedding of the weak gauge currents, we see $v^2 = N_D f^2$, where $f$ is the
technipion decay constant.

One may rewrite $S$ as an integral over a spectral function
\eqn\SdefIII{
S = {1 \over 3 \pi} \int_0^\infty {ds \over s}
\left\{\left[R_V(s)-R_A(s)\right] - {1\over 4}\left[1 - \left(1 -
m_H^2\right)^3 \theta(s - m_H^2) \right]\right\}~,
}
where $m_H$ is the reference value of the Higgs boson mass and
\eqn\Rdef{
\eqalign{
R_V(s) =& -12 \pi \hbox{Im }{\Pi_{VV}(s)\over s}\cr
R_A(s) =& -12 \pi \hbox{Im }{\Pi_{AA}(s)-\Pi_{AA}(0)\over s}~.
}
}
The second term in \SdefIII\ subtracts off the standard model contribution and
renders the expression finite.

The integral in \SdefIII\ is very convergent in the ultraviolet, and therefore
it is most sensitive to the behavior of the $R$'s in the infrared.  A simple
model for the functions $R_V$ and $R_A$ is that they are each concentrated at
their lightest resonance\foot{This model has no infrared problem, so we
neglect the weak dependence on $m_H$ in the following discussion.}:
\eqn\vecdom{
\eqalign{
R_V(s) =& 12 \pi^2 F_{\rho T} \delta(s-m_{\rho T}^2) \cr
R_A(s) =& 12 \pi^2 F_{a_1 T} \delta(s-m_{a_1 T}^2)~,
}
}
where the $F$s have dimensions of mass, and $m_{\rho T}$ and $m_{a_1 T}$ are
the techni-$\rho$ and techni-$a_1$ masses respectively.  The first and second
Weinberg sum rules \ref\Wsum{C.~Bernard \etal, \prd{12}{1976}{792}.}
imply relationships among these quantities
\ref\Wrel{S.~Weinberg, \prl{18}{1967}{507}.}
\eqn\Wrelat{
\eqalign{
F_{\rho T} = {m_{a_1 T}^2 N_D f^2 \over m_{a_1 T}^2 - m_{\rho T}^2} \cr
F_{a_1 T} = {m_{\rho T}^2 N_D f^2 \over m_{a_1 T}^2 - m_{\rho T}^2} ~.
}
}
One therefore obtains
\eqn\Sspec{
S = 4 \pi \left(1 + {m_{\rho T}^2 \over m_{a_1 T}^2} \right) {N_D f^2
\over m_{\rho T}^2}
}
In going from QCD to technicolor holding $N_{TC}$ fixed at 3, one scales the
masses of the vector technimesons so that $f/m$ is fixed \ref\DRK{S.
Dimopoulos, S. Raby, and G. Kane, \np {B182} {1981} {77}}, and so $S$ is
proportional to one power of $N_D$.  To go to some other value of $N_{TC}$ one
uses the large-$N_c$ QCD \ref\largeN{G. 't Hooft, \np {B72} {1974} {461}.}
result that the ratio $f/m$ scales like $\sqrt{N_{TC}}$.  Plugging in the
observed masses of the $\rho$ and $a_1$, one obtains the result
\eqn\Sspecres{
S \sim 0.083 N_D N_{TC}
}
A more elaborate analysis which includes the width of the vector technimesons
gives roughly the same answer and has the same dependence on $N_D$ and
$N_{TC}$.

Next we turn to the computation in the chiral Lagrangian.  The two-point
function $\Pi_{3Y}$ has two contributions.  At tree level, the relevant number
is the coefficient\foot{This is related to the coefficient called $\ell_{10}$
in the notation of Gasser and Leutwyler \ref\GL{Gasser and Leutwyler, Ann.
Phys. (NY) {\bf 158} (1984) 142 and \np {B250} {1985} {465}.}. They also point
out that in the large-$N_c$ limit, $\ell_{10}$ is proportional to $N_c$.} $c$
of the operator $W_3^{\mu\nu} B_{\mu\nu}$, where $W_i^{\mu\nu}$ and
$B^{\mu\nu}$ are the $SU(2)_W$ and $U(1)_Y$ field strengths respectively. The
definition is
\eqn\SdefIV{
S = -32 \pi c~.
}
There are additional, formally infinite, contributions to $\Pi_{3Y}$
arising from loops of technipions. These divergences may be absorbed into a
renormalization of $c$ in the usual way, and it is the sum of the loop
contributions to $\Pi_{3Y}$ and that from $c$ that is finite. Because of the
three massless exact Goldstone bosons which are ``eaten'' by the $W$ and $Z$,
there is an infrared logarithmic divergence. As above however, the same
divergence exists in the standard model, and the same subtraction renders $S$
finite.

Using naive dimensional analysis we bound $S$ as follows.  One calculates the
graphs consisting of a loop of technipions, which induce a running of $c$.
The technipions contribute to the running from a high scale $\Lambda_\chi$, at
which the chiral Lagrangian breaks down, down to their mass, at which point
they are integrated out of the theory. Without a fine tuning, it is
inconsistent to assume that $S$ is smaller than this logarithmic contribution.

The technipions form (approximately) degenerate multiplets under the custodial
$SU(2)_C$; each one makes a contribution to $S$ of \GR \eqn\SpiI{ \Delta S =
{1 \over 36 \pi} \ell (\ell+1) (2\ell+1) F
\log\left({\Lambda_\chi^2 \over m_\pi^2}\right)~,
}
where $\ell$ is the isospin of the multiplet under the $SU(2)_c$, and $F$ is a
symmetry factor which is 1 for a non-self conjugate multiplet (like the $K^+$,
$K^0$ of QCD), and 1/2 for a self-conjugate multiplet (like $\pi^+$, $\pi^0$,
$\pi^-$).  The models we are considering have $N_D^2-1$ self-conjugate
triplets ($\ell = 1$) of massive technipions, $N_D N_S$ non-self-conjugate
doublets ($\ell = 1/2$), $N_D^2+N_S^2-1$ singlets ($\ell = 0$), and one
triplet of ``eaten'' technipions.  The contribution of this last set is
partially cancelled by the standard model subtraction, leaving a term
proportional to $\log(\Lambda_\chi^2/m_H^2)$, which we neglect.
Therefore we may write
\eqn\SpiII{ S \ge {1 \over 24 \pi} (N N_D - 2)
\log\left({\Lambda_\chi^2 \over m_\pi^2}\right)~, }

At this stage the large scale $\Lambda_\chi$ is arbitrary, but in an $SU(N)_L
\times SU(N)_R \times U(1) \to SU(N)_V \times U(1)$ chiral lagrangian it
cannot be larger\foot{In ref \GR, the $\Lambda_\chi$ was taken to be $4 \pi
f$. This is too large a scale.} than of order $4\pi f / \sqrt{N}$
\ref\SS{M.~Soldate and R.~Sundrum, \np {B340} {1990} {1}.}. Consider the
case of the one-family model ($N=8$, $N_D=4$). If we assume that this bound
is saturated then $\Lambda_\chi$ is about 550 GeV. If we take all the
technipions to have a mass of about 100 GeV, then $S$ is bigger than or
about 1, just as it was when it was evaluated using the spectral integrals
in the vector dominance model.

As was stressed in the introduction, this numerical coincidence is rather
surprising.  The chiral Lagrangian calculation displays no dependence on
$N_{TC}$, but goes as $N_D N - 2$ (ignoring the weak dependence in the
logarithm).  The vector dominance model computation went like $N_D N_{TC}$.
Certainly the two computations will be very different if we go to the
two-family model, for example, or even add some extra singlets.

We may explain the discrepancy by considering the graphs which the two
calculations have included.  The vector dominance model scaled with
large-$N_C$ QCD includes the class of diagrams denoted in \fig\vd{The class of
diagrams that is included by the large-$N_c$ vector dominance approximation.
Gluon lines are omitted.}, which goes as $N N_{TC}$.  The class of diagrams
included in the chiral Lagrangian is shown in \fig\chil{The class of diagrams
that is included in the chiral Lagrangian computation. Gluon lines are
omitted.} - proportional to $N N_D$.  Were we able to compute $S$ in the full
TC theory, both classes would be present, the outstanding question is which
one dominates.

\newsec{Kaplan's Argument}

Imagine that it is possible to solve exactly an $SU(N_{TC})$ gauge theory with
$N$ flavors of fermions.  We assume that the fermions are confined, and that
the observed spectrum consists entirely of TC singlets.  The $SU(N)_L \times
SU(N)_R \times U(1)$ chiral symmetry breaks to $SU(N)_V \times U(1)$ through
the formation of a condensate.  Aside from the $N^2-1$ technipions, all the
other particles are massive.  Make a plot of the mass $M$ of the lightest
massive resonance divided by the pion decay constant.  As $N_{TC}$ goes to
infinity with $N$ fixed, we know that this ratio goes as $1/\sqrt{N_{TC}}$
\largeN.  So if we know $M$ for a theory in which $N_{TC}$ is large, as we
reduce $N_{TC}$ the value of $M/f$ increases.

On the other hand, as argued in \ref\CDG{R.~S.~Chivukula, M.~Dugan, and
M.~Golden, ``Analyticity, Crossing Symmetry, and the Limits of Chiral
Perturbation Theory'', Boston University preprint BUHEP-92-18 (June 1992).},
there is an upper bound to $M/f$.  In a lowest-order chiral Lagrangian
computation of the $\pi \pi \to \pi \pi$ scattering process, the amplitude for
the $SU(N)_V$ singlet spin-0 partial wave is given by \ref\cs{R.~N.~Cahn and
M.~Suzuki, \prl {67} {1991} {169}.}
\eqn\wave{
a = {N s \over 32 \pi f^2}
}
where $s$ is the usual Mandelstam variable.  A partial wave amplitude must lie
on or inside the Argand circle, so when $s$ is greater than or about $4 \pi f
/ \sqrt{N}$, the corrections must be bigger than the lowest order computation,
indicating the likely divergence of the chiral Lagrangian's expansion of
amplitudes as a power series in energy.  This is the probable scale for the
formation of non-analytic structure in the $S$-matrix, such as resonances.  It
is unlikely that it is possible to postpone such structures much beyond this
mass, though they may be lighter.
\eqn\bound{
{M \over f} \le {4 \pi \over \sqrt{N}}
}
Once the mass of the lightest resonance saturates this bound, decreasing
$N_{TC}$ cannot increase $M/f$.

A plot of $M/f$ as a function of $N_{TC}$ might look something like one of the
lines in \fig\Mf{Hypothetical graphs of the mass of the lightest resonance
divided by the pion decay constant in an $SU(N_{TC})$ gauge theory with a
fixed number $N$ of fermions as a function of $N_{TC}$.}.  Here we imagine
that $N$ is fixed at some value. The solid line shows one possibility. At
large $N_{TC}$, the curve goes as $1/\sqrt{N_{TC}}$. At somewhat lower
$N_{TC}$, however, there is a flat part where the mass of the lightest
resonance saturates the bound \bound\ and changing $N_{TC}$ cannot much affect
$M/f$. At very low $N_{TC}$, the theory loses asymptotic freedom and ceases to
make sense. Another possibility is shown as the dashed line. Here $M/f$ is
always on the falling part of the curve, and the bound \bound\ is never
saturated.

In QCD as we know it, with two light flavors\foot{One may wish to argue
that there are three light quarks in QCD. Since the bound is only a rough
guide, it does not distinguish between $N=2$ and $N=3$.} and $N_c=3$, the
scale $4 \pi f / \sqrt{N}$ is about 825 MeV. The lightest resonances, such
as the $\rho(770)$, have masses of about this value and appear therefore to
be saturating the bound. Therefore, {\it in QCD as we know it, we are on a
curve like the solid one, and $N_c=3$ corresponds to a point like A or B,
not C!} If QCD is at a point like A, one has to increase $N_c$ well beyond
3 before the $1/\sqrt{N_c}$ dependence sets in.

\newsec{Application to Electroweak Corrections}

In the one-family model $4 \pi f / \sqrt{N}$ is about 550 GeV, and we
expect resonances of this mass or lower. On the other hand, the vector
dominance computation assumed that the technicolor spectrum was directly
analogous to the QCD spectrum and that (for $N_{TC}=3$) the ratio $m_{\rho
T}/f$ was the same as in QCD, yielding $m_{\rho T}=1$ TeV. Such a large
mass for the lightest resonance is inconsistent. The lightest resonance
(which may or may not be a vector meson) must be {\it lighter} than the
simple scaling suggests.

Suppose that we continue to make the (entirely unwarranted) assumption that
the spectrum of technicolor looks just like that of QCD, \ie\ we model the
$VV$ and $AA$ spectral functions at small $q^2$ as each being dominated by a
single spin-1 resonance.  However, instead of scaling $M/f$ from QCD, assume
that we are at a point like A or B in \Mf, and put $m_{\rho T} \approx 4 \pi
f / \sqrt{N}$.  Evaluating eqn. \Sspec\ we find:
\eqn\Sspecnew{
S = {1 \over 4 \pi} \left(1 + {m_{\rho T}^2 \over m_{a_1 T}^2} \right) N
N_D
}
The dependence of $S$ has lost one factor of $N_{TC}$, and gained a
factor of $N$.  In other words, the dependence on $N$, $N_D$, and $N_{TC}$ is
the same as in the chiral Lagrangian calculation using NDA!

Another possibility is that $N_{TC}$ is sufficiently large that we are at a
point like C, or that technicolor is on the dashed curve.  In this case, the
vector dominance assumption gives a dependence like $N_D N_{TC}$.  However, if
the techni-$\rho$ is the lightest resonance, its mass is less than 550 GeV,
and the scaling of $M/f$ from QCD is invalid.

\newsec{Double Scaling Limit}

In QCD neither $N$ nor $N_c$ is particularly large, and it is not clear
that the large $N_c$ approximation is particularly good.  Moreover, the
ratio $N/N_c$ is not small, and it may not be possible to neglect it.  In one
family technicolor this problem is exacerbated, since there are eight light
fermions, rather than two.

Consider instead the limit of QCD in which $N_c$ and $N$ are both taken to
infinity with their ratio held fixed \ref\Ven{G. Veneziano,
\np{B117}{1976}{519}\semi G. Veneziano, in {\it Proceedings of the Twelfth
Rencontre de Moriond, Flaine-haute-Savoie (France)}, Vol. III, Tran Thanh Van,
ed., 1977, p 113-134}.  The graphs that contribute now are planar, but with
holes for fermion loops.  The double scaling limit is a better approximation
to QCD than the ordinary large-$N_c$ QCD limit, in the sense that all the
diagrams leading in $N_c$ are included, plus some extra ones. If one believes
that QCD is well approximated by its large-$N_c$ limit, then the double
scaling limit is also justified.

In this double limit, \vd\ and \chil\ are both the same order, so it is not
possible to know which one is larger.  Both \Sspecres\ and \SpiII\ have the
same dependence.  In the double scaling limit, the ratios of the masses of the
resonances to $f$ falls like $1/\sqrt{N_{TC}}$ as before, but the ratio of the
masses to the widths stays fixed.  This is because the number of open channels
into which the resonances decay grows with $N$.

If we believe that QCD with $N_c=3$ and $N=2$ is close to the double
scaling limit, then the one-family model most closely resembles QCD at
$N_{TC} = 12$. Therefore (in the double scaling limit) when $N_{TC}=12$,
the one family model is at A or B, not C. Since QCD saturated the bound on
$M/f$, all one-family models with $N_{TC} \le 12$ must also saturate the
bound.

For the one-family model with $N_{TC}=12$, \Sspecnew\ yields\foot{Recall
that NDA gives a {\it lower bound} on $S$, and this estimate is consistent
with it.} $S \approx 4$. If vector dominance continues to work when
$N_{TC}$ is less than 12, and if the lightest resonance is still the
techni-$\rho$, and if it continues to saturate the bound on its mass, then
reducing $N_{TC}$ will not reduce the value of $S$. The electroweak
corrections in one-family technicolor could be far larger than was
previously estimated.

On the other hand a great deal of caution should be advised. The use of vector
dominance in a model with a radically different, larger, value of $N/N_{TC}$
is highly speculative. Kaplan's argument discusses only the scale at which the
lightest resonance forms -- not the quantum numbers of the lightest resonance
or the behavior of the theory at higher energies.  There is no reason to
believe (as \Sspecnew\ seems to imply) that all $SU(N_{TC})$ gauge theories
with $N_{TC}$ in the flat part of the curve are identical, or even similar.
For example, as suggested by Cahn and Suzuki, the lightest resonance could be
a scalar instead of a vector particle \cs\ref\cg{ R.~S.~Chivukula and
M.~Golden, Nucl. Phys. {\bf B372} (1992) 44.}.  It is possible that the
scalar's appearance delays the formation of the techni-$\rho$ to a somewhat
higher scale.  We really do not know very much about the spectrum of
technicolor theories.

Lastly, the masses of the technipions may be very different from the scaled
value of the pion mass.  Certainly this will affect the size of the chiral
logs.  More importantly, if the technipions are sufficiently heavy, they
decouple from the low-energy amplitudes, effectively reducing the number of
flavors and increasing the bound on the mass of the lightest resonance.  This
is why the $\rho$ has a mass of 770 MeV, well above the 475 MeV bound on its
mass it would have if all six quarks were light.  It may be possible to
construct a technicolor model which avoids large radiative corrections if the
technipions are heavy enough.

\newsec{Conclusions}

Kaplan's argument addresses a longstanding question: How are NDA and the
large-$N_c$ expansion consistent? The diagrams used in NDA are always
subleading in a large-$N_c$ expansion, and yet dimensional analysis generally
gives reasonable values for the sizes of coefficients of the higher dimension
operators in the chiral Lagrangian of QCD. The arguments given here show that
NDA will work to estimate the sizes of these coefficients for values of $N_c$
such that the theory is on the flat part of the curve in \Mf, at a point like
A or B. It appears that ordinary QCD with $N_c=3$ is such a theory.

We have seen that there is a simple way to reconcile the vector dominance
and chiral Lagrangian/NDA calculations of $S$. If, as we have argued,
technicolor and QCD are on the flat part of the curve in \Mf, then the
dependence on $N$ and $N_D$ suggested by the chiral Lagrangian is
appropriate. If, on the other hand, a technicolor model has values of
$N_{TC}$ which makes the masses of the mesons much lighter than $4 \pi f /
\sqrt{N}$, then that theory is on the falling part of the curve, and
scaling the radiative corrections with $N_{TC}$ may be justified. In either
case, taking $M/f$ from QCD is invalid.

At first glance, it seems that the radiative corrections in one-family
technicolor may be much larger than previously estimated.  On the other hand,
we prefer to advise caution, since technicolor is not just QCD scaled up.  It
is a much less familiar theory than was assumed.

\bigskip

\noindent {\bf Acknowledgements. }\hfil\break

We thank Howard Georgi, David Kaplan, Aneesh Manohar, and Michael Peskin
for useful conversations. R.S.C. acknowledges the support of a NSF
Presidential Young Investigator Award, an Alfred P. Sloan Foundation
Fellowship, and a DOE Outstanding Junior Investigator Award and thanks the
Aspen Center for Physics for its hospitality during the completion of this
work. M.G. and R.S.C. thank the Texas National Research Laboratory
Commission for support under Superconducting Super Collider National
Fellowships. This research is supported in part by the National Science
Foundation under grants PHY--87--14654 and PHY--90--57173, by the Texas
National Research Laboratory Commission, under grants RGFY9206 and
RGFY92B6, and by the Department of Energy, under contract numbers
DE--AC02--89ER40509 and DE--FG02--91ER40676.

\listrefs
\listfigs

\input pictex
\nopagenumbers
$$
\beginpicture
\setcoordinatesystem units <1in,1in>
\setplotarea x from 0 to 4, y from 0 to 4
\ellipticalarc axes ratio 2:1 360 degrees from 3 2 center at 2 2
\normalgraphs
\plot
0.95 1.95
1.05 2.05
/
\plot
0.95 2.05
1.05 1.95
/
\plot
2.95 1.95
3.05 2.05
/
\plot
2.95 2.05
3.05 1.95
/
\plot
1.975 2.50
2.025 2.53
/
\plot
1.975 2.50
2.025 2.47
/
\plot
2.025 1.50
1.975 1.53
/
\plot
2.025 1.50
1.975 1.47
/
\endpicture
$$
\vfill\eject
$$
\beginpicture
\setcoordinatesystem units <1in,1in>
\setplotarea x from 0 to 4, y from 0 to 4
\ellipticalarc axes ratio 2:1 360 degrees from 3 2 center at 2 2
\ellipticalarc axes ratio 2:1 360 degrees from 2.6 2 center at 2 2
\normalgraphs
\plot
0.95 1.95
1.05 2.05
/
\plot
0.95 2.05
1.05 1.95
/
\plot
2.95 1.95
3.05 2.05
/
\plot
2.95 2.05
3.05 1.95
/
\plot
2.025 2.30
1.975 2.33
/
\plot
2.025 2.30
1.975 2.27
/
\plot
1.975 1.70
2.025 1.73
/
\plot
1.975 1.70
2.025 1.67
/
\plot
1.975 2.50
2.025 2.53
/
\plot
1.975 2.50
2.025 2.47
/
\plot
2.025 1.50
1.975 1.53
/
\plot
2.025 1.50
1.975 1.47
/
\endpicture
$$
\vfill\eject
$$
\beginpicture
\setcoordinatesystem units <0.0625in,2.4in>
\setplotarea x from 0 to 64, y from 0 to 1.25
\axis bottom visible /
\axis left visible /
\put {${\displaystyle 4 \pi \over \sqrt{N}}$} at -4 1
\putrule from -1 1 to 0 1
\put {$\bullet$} at 6 1
\put {$\bullet$} at 25 .95
\put {$\bullet$} at 49 .714
\put {A} at 6 1.05
\put {B} at 25 1.00
\put {C} at 49 .764
\put {$N_{TC}$} at 50 -0.06
\put {$M/f$} at -5 .7
\normalgraphs
\plot
3 1
9 1
16 1
/
\plot
16 1
20 0.99
25 0.95
36 0.8333
42 0.772
49 0.714
52 0.693
59 0.651
64 0.625
/
\setdashes
\plot
3 0.7500
6 0.5303
9 0.4330
12 0.3750
15 0.3354
18 0.3062
21 0.2835
24 0.2652
27 0.2500
30 0.2372
33 0.2261
36 0.2165
39 0.2080
42 0.2004
45 0.1936
48 0.1875
51 0.1819
54 0.1768
57 0.1721
60 0.1677
63 0.1637
64 0.1624
/
\endpicture
$$
\bye